\documentclass[%
 reprint,
 superscriptaddress,
 amsmath,%amssymb,
 aps,
 prb
]{revtex4-2}

\usepackage[pdftex]{graphicx, color}% Include figure files
\usepackage[pdftex,
            colorlinks=true,
            citecolor=blue,
            linkcolor=blue]{hyperref}
\usepackage{newtxtext, newtxmath}
\usepackage{bm}% bold math
\usepackage{braket, mathtools}
\usepackage{tabularx}
\graphicspath{{./images/}}

\DeclareMathOperator{\sgn}{sgn}
\DeclareMathOperator{\tr}{tr}
\newcommand{\ii}{i}

\begin{document}

\title{Supercurrent-Induced Weyl Superconductivity}
%\thanks{A footnote to the article title}

\author{Shuntaro Sumita}
\email[]{shuntaro.sumita@riken.jp}
\affiliation{%
 Condensed Matter Theory Laboratory, RIKEN CPR, Wako, Saitama 351-0198, Japan
}%

\author{Kazuaki Takasan}
\affiliation{%
 Department of Physics, University of California, Berkeley, California 94720, USA
}%
\affiliation{%
 Materials Sciences Division, Lawrence Berkeley National Laboratory, Berkeley, CA 94720, USA
}%
\affiliation{%
 Department of Physics, University of Tokyo, 7-3-1 Hongo, Tokyo 113-0033, Japan
}%

\date{\today}

\begin{abstract}
 We show that Weyl superconductivity can be induced by finite supercurrent in noncentrosymmetric spin-orbit-coupled superconductors with line nodes.
 We introduce a three-dimensional tight-binding model of a tetragonal superconductor in a $D+p$-wave pairing state with a finite center-of-mass momentum, and elucidate that a line-nodal to point-nodal spectral transition occurs by applying an infinitesimal supercurrent.
 We also clarify that the higher-order effect in spin-orbit coupling is particularly important for this phenomenon.
 The point nodes are protected by topologically nontrivial Weyl charges, and therefore gapless arc states appear on the surface of the superconductor.
 Furthermore, both the positions and the Weyl charges of the point nodes depend on the direction of the current.
 In addition, a quantized Berry phase defined on high-symmetry planes characterizes the Weyl nodes when the in-plane supercurrent is considered.
 Our proposition paves a new way for controlling the superconducting gap structures by using an external field.
\end{abstract}

\maketitle

\section{Introduction}
A superconducting gap is one of the key parameters in the research of superconductivity.
In conventional superconductors described by Bardeen--Cooper--Schrieffer (BCS) theory in 1957~\cite{BCS}, the superconducting gap, namely excitation energy of the Bogoliubov spectrum, has a fully gapped $s$-wave structure.
Since 1980s, on the other hand, a lot of unconventional superconductors beyond the BCS theory hosting nodes in the gap structure have been discovered.
High-$T_c$ cuprates are representative examples of the nodal superconductors.
Although the cuprate superconductors were considered to have $s$-wave pairing symmetry in the early stage of the study, many experiments in 1990s have reported various evidence for the anisotropic $d_{x^2 - y^2}$-wave gap structure with line nodes~\cite{Harlingen1995_review, Tsuei2000_review, Hardy1993}.
For example, Hardy \textit{et al.} have found that the low-temperature magnetic penetration depth in YBa$_2$Cu$_3$O$_{7-\delta}$ is linear in temperature $T$~\cite{Hardy1993}, which is different from the exponential behavior in fully gapped superconductors.
Furthermore, the angle-resolved photoemission spectroscopy (ARPES) has directly observed the momentum dependence of the superconducting gap structure, where the line nodes emerge in the diagonal direction of $k_x \pm k_y = 0$~\cite{Shen1993, Shen1995, Ding1995, *[][{ (ettarum).}]{Ding1995_erratum}, Norman1995_BSCCO, Ding1996}.

As indicated in the above examples, the superconducting gap structure plays an important role in discussing the symmetry of the order parameter and the pairing mechanism.
From the theoretical point of view, previous studies have constructed classification methods for predicting the superconducting gap structure by using symmetry and topology~\cite{Volovik1984, Volovik1985, Anderson1984, Ueda1985, Blount1985, Sigrist-Ueda, Nomoto2016_PRB, Yarzhemsky1992, Yarzhemsky1998, Yarzhemsky2000, Yarzhemsky2003, Yarzhemsky2008, Yarzhemsky2018, Yarzhemsky2021, Norman1995_node, Micklitz2009, Micklitz2017_PRB, Micklitz2017_PRL, Yanase2016, Nomoto2017, Bzdusek2017, Kobayashi2014, Kobayashi2016, Kobayashi2018, Sumita2018, Sumita2019, Sumita2021_book, Samokhin2019, Ono2022, Tang2021_arXiv, Sumita2022_review}.
In particular, recent theoretical studies have revealed unconventional gap nodes owing to the characteristics of nonsymmorphic superconductors~\cite{Norman1995_node, Micklitz2009, Micklitz2017_PRB, Micklitz2017_PRL, Yanase2016, Nomoto2017, Kobayashi2016, Kobayashi2018, Sumita2018, Nomoto2016_PRL, Sumita2017} and multi-orbital ones~\cite{Nomoto2016_PRB, Sumita2018, Sumita2019, Samokhin2019, Brydon2016, Agterberg2017, Timm2017, Savary2017, Boettcher2018, Venderbos2018}, which could not be explained by the earlier classification theory of the superconducting order parameter based on point group symmetry~\cite{Volovik1984, Volovik1985, Anderson1984, Ueda1985, Blount1985, Sigrist-Ueda}.
For example, a heavy-fermion superconductor UPt$_3$ is considered to possess nontrivial line nodes protected by nonsymmorphic symmetry~\cite{Norman1995_node, Micklitz2009, Micklitz2017_PRB, Micklitz2017_PRL, Yanase2016, Kobayashi2016, Kobayashi2018, Sumita2018, Nomoto2016_PRL} and a lot of Weyl nodes in the time-reversal symmetry breaking B phase~\cite{Yanase2016, Sumita2019}, which are compatible with experimental observations~\cite{Sauls1994_review, Joynt2002_review}.
Also, in other superconductors, some of the theoretically predicted nontrivial gap structures have been reported in experiments~\cite{Badger2022, Kim2018}.
Therefore, the synergetic effect between experimental and theoretical studies has promoted the understanding of the nontrivial superconducting gap structures and the related pairing mechanisms.

It has been known that the superconducting gap structure can be predicted by the temperature dependence of various physical quantities, e.g., penetration depth, specific heat, thermal conductivity, and NMR relaxation rate~\cite{Sigrist-Ueda, Lapp2020}.
For experimental measurements of the observables, in many cases we need to apply an external field such as a magnetic field.
In this sense, the (weak) field is necessary to \textit{detect} the superconducting nodes, which are protected by symmetry as classified in the above theory.
On the other hand, the external field itself has the potential to break some of the symmetry and to change the nodal structure.
Through the modification of the excitation spectrum, we can use the external field to \textit{control} the quantum state.
Indeed, previous theoretical studies have suggested that fully gapped topological superconductivity can be realized by applying a Zeeman field~\cite{Daido2016, Daido2017} or laser light~\cite{Takasan2017} to noncentrosymmetric $d$-wave superconductors.
Furthermore, in our previous work, we have elucidated the possibility of supercurrent-induced topological phase transition in two-dimensional (2D) noncentrosymmetric superconductors~\cite{Takasan2021_arXiv}.
Such a topological superconducting phase with a finite center-of-mass (COM) momentum of the Cooper pairs has garnered significant attention recently~\cite{Romito2012, Lesser2021, Zhang2013, Qu2013, Nissinen2017, Volovik2018, Ying2018, Ying2019, Hu2019, Dmytruk2019, Papaj2021, Volkov2020_arXiv}.

\begin{figure}[tbp]
 \centering
 \includegraphics[width=\linewidth, pagebox=artbox]{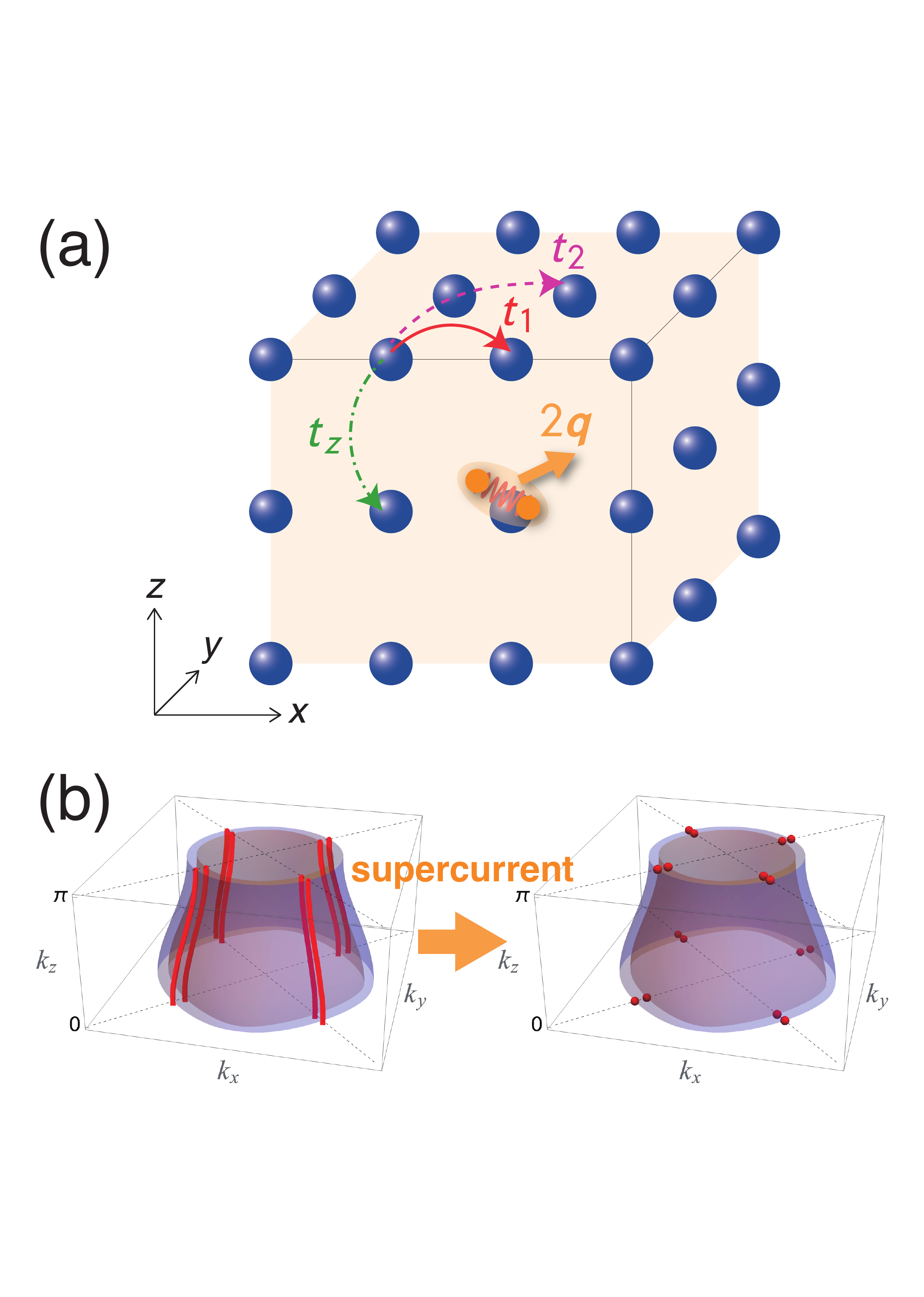}
 \caption{(Color online) (a) A schematic picture of our model on the tetragonal lattice (the blue solid circles). The parameters $t_1$ and $t_2$ correspond to in-plane nearest- and next-nearest-neighbor hopping integrals, respectively (the red solid and purple dashed arrows), while $t_z$ is a nearest-neighbor hopping along the $z$ direction (the green dashed-dotted arrow). Supercurrent is carried by Cooper pairs with the COM momentum $2\bm{q}$. (b) The change of superconducting node structures discussed in this paper. Line nodes (the red solid lines in the left figure) are partially gapped out to be point nodes (the red points in the right figure) under the finite supercurrent.}
 \label{fig:system_and_nodes}
\end{figure}

Stimulated by the backgrounds explained above, in this paper we propose that Weyl superconductivity~\cite{Meng2012, *[][{ (ettarum).}]{Meng2012_erratum}} can be realized by applying supercurrent to three-dimensional (3D) noncentrosymmetric line-nodal superconductors with spin-orbit coupling.
The basic concept of the study is schematically shown in Figs.~\ref{fig:system_and_nodes}(a) and \ref{fig:system_and_nodes}(b).
We consider a 3D model of a $D+p$-wave superconductor with a noncentrosymmetric tetragonal lattice under a finite supercurrent, where the Cooper pairs have a COM momentum $2\bm{q}$ [Fig.~\ref{fig:system_and_nodes}(a)].
As illustrated in Fig.~\ref{fig:system_and_nodes}(b), the line-nodal structure in the $D+p$-wave superconductor is modified into a point-nodal one by the supercurrent, in which each point node is characterized by a nontrivial Weyl charge~%
\footnote{Strictly speaking, the word ``node'' here is used in a different meaning from its normal sense. In the usual sense, where a node means a gapless point at the \textit{zero} energy, there exist surface nodes (Bogoliubov Fermi surfaces) under the finite supercurrent because of the Doppler shift~\cite{Matsuda2006}. However, our main concern in this paper is how the particle-like energy bands touch the hole-like bands, generally at a \textit{finite} energy value.}.
We find that the higher-order term of the spin-orbit coupling plays a crucial role in the supercurrent-induced Weyl superconductivity.
Notably, our calculations demonstrate that the positions and Weyl charges of the point nodes depend on the direction of the supercurrent.
Little is known about such a controlling way of the Weyl nodes in superconductors by an external field, besides only a few reports on tunable positions of Weyl points in UPt$_3$ by a magnetic field and temperature~\cite{Yanase2016, Sumita2019}.
Moreover, when the direction of the supercurrent is in-plane, the Weyl nodes must be fixed on $k_z = 0$ and $k_z = \pi$ because of the existence of an additional topological index, namely a quantized Berry phase.

Note that the transition between a line-nodal structure and a point-nodal one with Weyl charges is already known to be induced in the A phase of superfluid ${}^3$He by cooling, pressure, or superflow~\cite{Nissinen2018, Makinen2019, Autti2020}.
However, to our knowledge, such a transformation of nodal structures has not been much studied in the context of superconductors.
Furthermore, one of the key ingredients for our suggestion is the spin-orbit coupling, which does not play a major role in the superfluid ${}^3$He.

The paper is organized as follows.
In Sect.~\ref{sec:model} our tight-binding model of the 3D spin-orbit-coupled $D+p$-wave superconductor under the supercurrent is presented.
In Sect.~\ref{sec:energy_spectrum} we obtain the Bogoliubov quasiparticle energy spectrum of the model, and show the change from a line-nodal to a point-nodal structure by applying supercurrent.
Next, we discuss the two topological numbers, i.e., the Weyl charge and the quantized Berry phase, both of which characterize the supercurrent-induced point nodes (Sect.~\ref{sec:topology}).
Surface arc states corresponding to the nontrivial Weyl charge are also shown.
Finally, a summary and discussion are given in Sect.~\ref{sec:summary}.

\section{Model}
\label{sec:model}
In this section, we introduce a 3D tight-binding model of an inversion-symmetry-breaking $D+p$-wave superconductor with finite supercurrent on a tetragonal lattice.
The $D+p$-wave superconductivity is not artificial since such a parity-mixed order is stabilized by an antiferromagnetic fluctuation in the Hubbard model~\cite{Yanase2007, Yanase2008}.
Since we are interested in clean superconductors without magnetic fields, the supercurrent is considered to be dominated by the current of the Cooper pairs.
In this study, we assume that all the Cooper pairs have the same COM momentum $2\bm{q}$.
A similar treatment is used in Refs.~\cite{Ikeda2020, Daido2022, Chazono2022, Takasan2021_arXiv}.
Under the above assumption, the mean-field Hamiltonian is given by
\begin{equation}
 H = \frac{1}{2} \sum_{\bm{k}} \bm{\Psi}_{\bm{k}; \bm{q}}^\dagger H_{\text{BdG}}(\bm{k}; \bm{q}) \bm{\Psi}_{\bm{k}; \bm{q}},
\end{equation}
where $\bm{\Psi}_{\bm{k}; \bm{q}} = (c_{\bm{k}+\bm{q}, \uparrow}, c_{\bm{k}+\bm{q}, \downarrow}, c_{-\bm{k}+\bm{q}, \uparrow}^\dagger, c_{-\bm{k}+\bm{q}, \downarrow}^\dagger)^{\text{T}}$ with $c_{\bm{k}, s}$ being the annihilation operator of electrons carrying momentum $\bm{k}$ and spin $s$.
The Bogoliubov--de Gennes (BdG) Hamiltonian matrix is represented by the following form,
\begin{equation}
 H_{\text{BdG}}(\bm{k}; \bm{q}) =
 \begin{bmatrix}
  H_{\text{N}}(\bm{k} + \bm{q}) & \Delta(\bm{k}) \\
  \Delta^\dagger(\bm{k}) & - H_{\text{N}}^{\text{T}}(-\bm{k} + \bm{q})
 \end{bmatrix},
 \label{eq:Hamiltonian}
\end{equation}
where
\begin{align}
 H_{\text{N}}(\bm{k}) &= \xi(\bm{k}) \sigma_0 + \bm{g}(\bm{k}) \cdot \bm{\sigma},
 \label{eq:normal_Hamiltonian} \\
 \Delta(\bm{k}) &= [\psi(\bm{k}) \sigma_0 + \bm{d}(\bm{k}) \cdot \bm{\sigma}] \ii\sigma_y,
 \label{eq:order_parameter}
\end{align}
are the normal-part Hamiltonian and the superconducting order parameter, respectively.
$\bm{\sigma} = (\sigma_x, \sigma_y, \sigma_z)$ represents the Pauli matrices in the spin space.

The kinetic energy term and the antisymmetric spin-orbit coupling (ASOC) term in Eq.~\eqref{eq:normal_Hamiltonian} are given by,
\begin{align}
 \xi(\bm{k}) &= - 2t_1 (\cos k_x + \cos k_y) - 4t_2 \cos k_x \cos k_y \notag \\
 & \quad - 2t_z \cos k_z - \mu,
 \label{eq:kinetic_term} \\
 \bm{g}(\bm{k}) &= \alpha_1 (- \sin k_y \hat{x} + \sin k_x \hat{y}) \notag \\
 & \quad + \alpha_2 \sin k_x \sin k_y \sin k_z (\cos k_x - \cos k_y) \hat{z},
 \label{eq:g-vector}
\end{align}
where $t_1$, $t_2$, $t_z$ are hopping parameters [see Fig.~\ref{fig:system_and_nodes}(a)], and $\mu$ is a chemical potential.
The first term in the $g$-vector [Eq.~\eqref{eq:g-vector}] represents a well-known Rashba-type ASOC, while the second term takes into account higher-order effects about the momentum $\bm{k}$.
The two terms have the same symmetry; strictly speaking, both $g$-vectors are classified into the irreducible representation $A_{2u}$ of the crystal point group $D_{4h}$~\cite{Frigeri_PhD}.
In particular, the higher-order term plays an essential role in the supercurrent-induced Weyl superconductivity (see Sect.~\ref{sec:energy_spectrum_in}).

The spin-singlet ($d$-wave) and spin-triplet ($p$-wave) components of the parity-mixed superconducting order parameter [Eq.~\eqref{eq:order_parameter}] are given by
\begin{align}
 \psi(\bm{k}) &= \Delta_d (\cos k_x - \cos k_y),
 \label{eq:order_parameter_d} \\
 \bm{d}(\bm{k}) &= \Delta_p (\sin k_y \hat{x} + \sin k_x \hat{y}),
 \label{eq:order_parameter_p}
\end{align}
respectively.
Both order parameters belong to the irreducible representation $B_1$ of the point group $C_{4v}$~\cite{Sigrist-Ueda}.

In the following sections, we analyze the Hamiltonian and show the appearance of supercurrent-induced Weyl superconductivity.
Throughout this paper, the model parameters are chosen to be 
\begin{align}
 & (t_1, t_2, t_z, \alpha_1, \alpha_2, \mu, \Delta_d, \Delta_p) \notag \\
 &= (1, 0.2, 0.5, 0.3, 0.3, -0.7, 0.5, 0.2),
 \label{eq:parameter}
\end{align} 
unless explicitly mentioned otherwise.

\section{Quasiparticle Energy Spectrum}
\label{sec:energy_spectrum}

\begin{figure*}[tbp]
 \centering
 \includegraphics[width=\linewidth, pagebox=artbox]{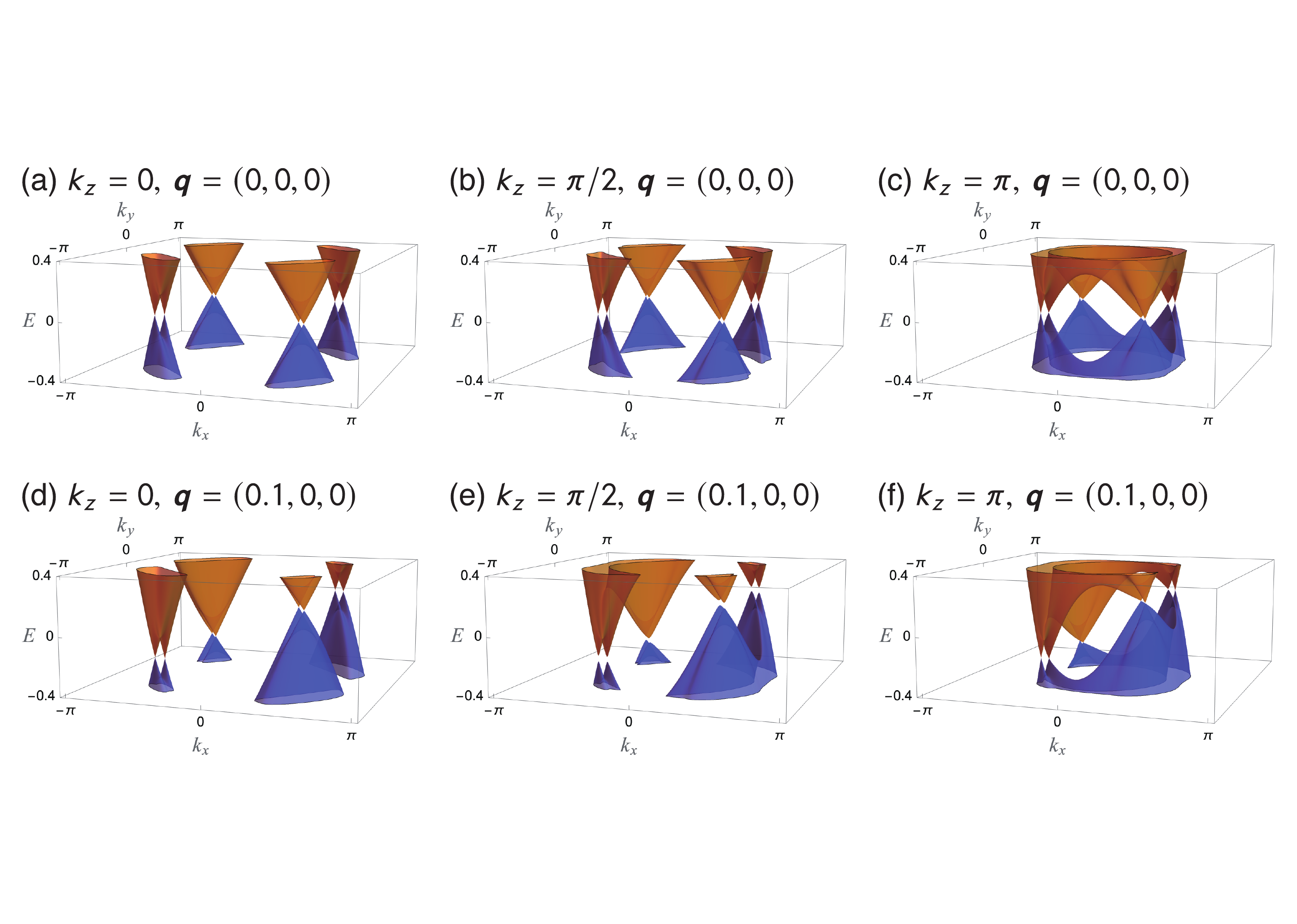}
 \caption{(Color online) Quasiparticle energy spectra on (a, d) $k_z = 0$, (b, e) $k_z = \pi/2$, and (c, f) $k_z = \pi$ planes. The COM momentum $\bm{q}$ of the Cooper pairs is set to be zero in the upper panels (a--c), while it is $(0.1, 0, 0)$ in the lower panels (d--f).}
 \label{fig:spectrum_slice_1}
\end{figure*}

In this section, we show the energy spectrum of Bogoliubov quasiparticles by diagonalizing the BdG Hamiltonian [Eq.~\eqref{eq:Hamiltonian}].
Figures~\ref{fig:spectrum_slice_1}(a)--\ref{fig:spectrum_slice_1}(c) represent the $(k_x, k_y)$-dependence of the energy spectra on $k_z = 0$, $\pi/2$, and $\pi$, respectively, in the absence of the supercurrent.
All the spectra have gapless points along the $[110]$ and $[1{-1}0]$ directions, since there exist line nodes on $k_x \pm k_y = 0$ due to the $d_{x^2-y^2}$-wave order parameter [Eq.~\eqref{eq:order_parameter_d}].
The line nodes are protected by symmetry and topology; indeed, chiral symmetry $\Gamma$ is well-defined for $\bm{q} = \bm{0}$, and a one-dimensional (1D) winding number defined on a loop $C$ encircling the node,
\begin{equation}
 w_C = - \frac{1}{4\pi i} \oint_{C} \mathrm{d}\bm{k} \cdot \tr[\Gamma H_{\text{BdG}}^{-1}(\bm{k}; \bm{0}) \nabla_{\bm{k}} H_{\text{BdG}}(\bm{k}; \bm{0})],
 \label{eq:winding_num}
\end{equation}
takes a nontrivial value~\cite{Sato2011, Kobayashi2014, Kobayashi2018}.
Furthermore, the gapless points are also characterized by a zero-dimensional (0D) topological invariant~\cite{Kobayashi2014, Kobayashi2018}.
Since mirror symmetry $M_{[110]}$ ($M_{[1{-1}0]}$) is preserved on the $k_x + k_y = 0$ ($k_x - k_y = 0$) plane, the crossing on the gapless points is explained by the difference of occupation numbers in between the $M_{[110]} = +\ii$ and $M_{[110]} = -\ii$ ($M_{[1{-1}0]} = +\ii$ and $M_{[1{-1}0]} = -\ii$) sectors.

In the following, let us consider how the line-nodal superconducting gap structure is changed by applying finite supercurrent.

\subsection{The case of in-plane supercurrent}
\label{sec:energy_spectrum_in}
First, we show the results when the direction of the COM momentum is in the $x$--$y$ plane, namely, $q_z = 0$.
The quasiparticle energy spectra for $\bm{q} \parallel \hat{x}$ on $k_z = 0$, $\pi/2$, and $\pi$ are represented in Figs.~\ref{fig:spectrum_slice_1}(d)--\ref{fig:spectrum_slice_1}(f), respectively.
Comparing them with Figs.~\ref{fig:spectrum_slice_1}(a)--\ref{fig:spectrum_slice_1}(c), we can find that the gapless points on $k_z = \pi/2$ is opened by the finite supercurrent, while those on $k_z = 0, \pi$ remain.
This indicate that the \textit{line-nodal} structure of the $D+p$-wave superconductivity is changed into the \textit{point-nodal} one.
Indeed, the nodes under the finite supercurrent are Weyl points characterized by a nonzero Chern number; for detailed discussion, see Sect.~\ref{sec:topology_Weyl}.
As we confirm in Sect.~\ref{sec:topology_Berry}, furthermore, the point nodes must be fixed on the $k_z = 0, \pi$ planes when the supercurrent is restricted to the in-plane direction.

\begin{figure}[tbp]
 \centering
 \includegraphics[width=\linewidth, pagebox=artbox]{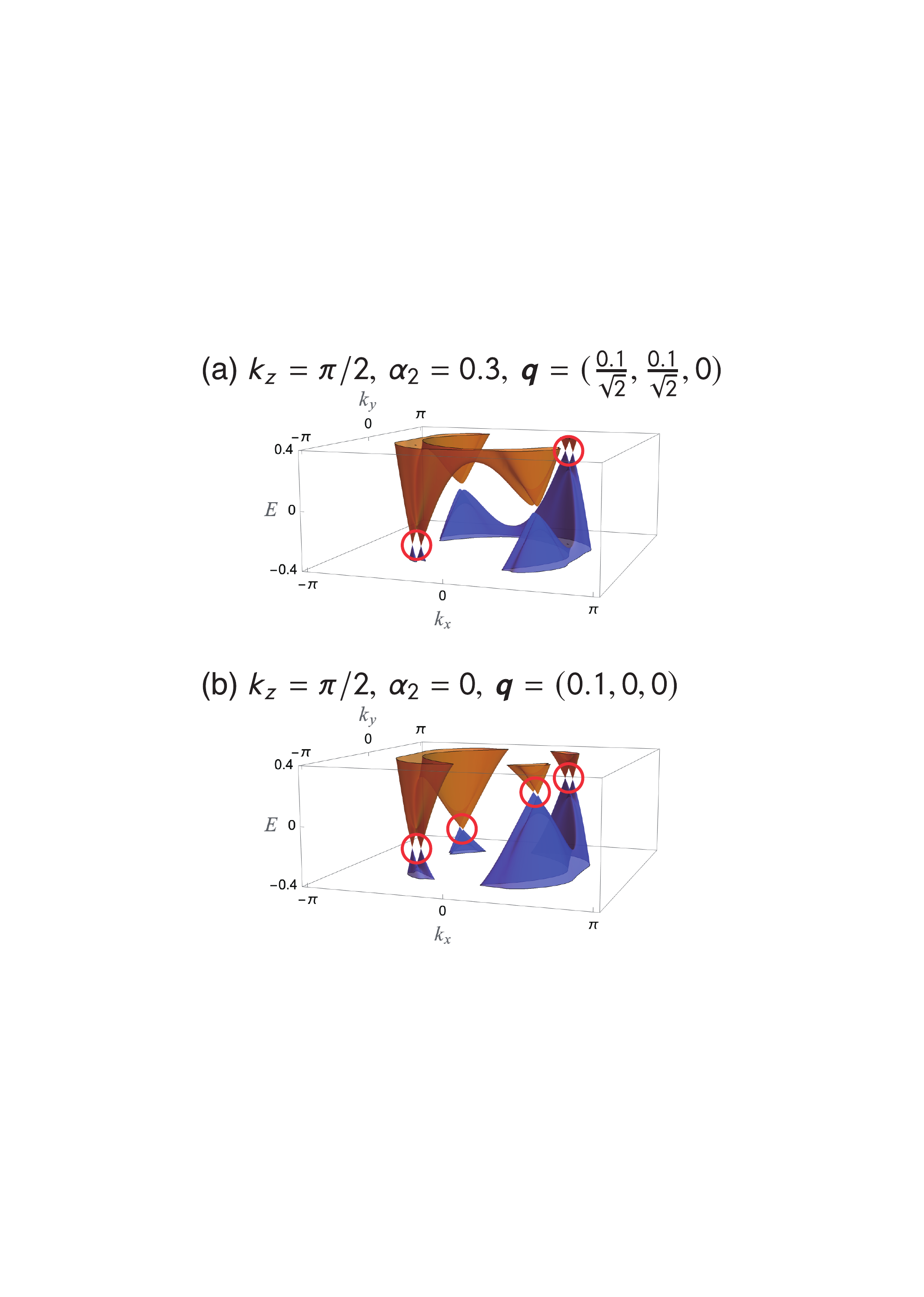}
 \caption{(Color online) Quasiparticle energy spectra on the $k_z = \pi/2$ plane for (a) $\alpha_2 = 0.3, \bm{q} = (\frac{0.1}{\sqrt{2}}, \frac{0.1}{\sqrt{2}}, 0)$ and (b) $\alpha_2 = 0, \bm{q} = (0.1, 0, 0)$. Unlike Fig.~\ref{fig:spectrum_slice_1}(e), gap opening does not occur under the supercurrent for (a) the [110] direction and (b) all directions (the red open circles).}
 \label{fig:spectrum_slice_2}
\end{figure}

Even when we consider a general in-plane direction of the supercurrent, the change in the quasiparticle energy spectrum from a line-nodal structure to a point-nodal one occurs, just like the above results.
However, only the $[110]$ (or $[1{-1}0]$) direction is special.
Figure~\ref{fig:spectrum_slice_2}(a) shows the energy spectrum with the $[110]$-directed COM momentum.
Obviously, the gapless points on $k_x - k_y = 0$ remain despite the finite supercurrent [see the red open circles in Fig.~\ref{fig:spectrum_slice_2}(a)].
The reason is that the mirror symmetry $M_{[1{-1}0]}$ is preserved in the special case.
As we discussed at the beginning of Sect.~\ref{sec:energy_spectrum}, line nodes in the zero-current state are protected by the mirror symmetry as well as the 1D winding number.
When we apply the supercurrent along the $[110]$ direction, one of the mirror symmetries $M_{[1{-1}0]}$ is preserved, while the other $M_{[110]}$ and the winding number are ill-defined.
Therefore, the line nodes on $k_x - k_y = 0$ remain, whereas those on $k_x + k_y = 0$ are changed to point nodes.

Furthermore, we emphasize the importance of the higher-order term in the $g$-vector [Eq.~\eqref{eq:g-vector}] for the occurrence of the supercurrent-induced Weyl nodes.
Indeed, the line nodes do not disappear under the supercurrent when $\alpha_2$ is zero [Fig.~\ref{fig:spectrum_slice_2}(b)].
The robustness of the line nodes can be understood by perturbation theory discussed in Ref.~\cite{Takasan2021_arXiv}.
Supposing that the applied supercurrent is small, the magnitude of the current-induced energy gap at the original nodal points $\bm{k}_0$ is estimated by
\begin{equation}
 \Delta_{\text{node}}(\bm{k}_0; \bm{q}) = 2 \frac{\left| \bm{g}'_{\bm{q}}(\bm{k}_0) \cdot [\bm{g}(\bm{k}_0) \times \bm{d}(\bm{k}_0)] \right|}{\left| \bm{g}(\bm{k}_0) \right|^2} + \mathcal{O}(q^2),
 \label{eq:gap_by_current}
\end{equation}
where $\bm{g}'_{\bm{q}}(\bm{k}) = (\bm{q} \cdot \nabla_{\bm{k}}) \bm{g}(\bm{k})$.
Equation~\eqref{eq:gap_by_current} indicates that the mixed spin-triplet $p$-wave component, namely the $d$-vector, is important for the current-induced gap.
Moreover, the gap size must be zero when $\alpha_2 = 0$, since all the vectors $\bm{g}'_{\bm{q}}(\bm{k})$, $\bm{g}(\bm{k})$, and $\bm{d}(\bm{k})$ are in the $x$--$y$ plane.
When the higher-order term in Eq.~\eqref{eq:g-vector} is finite, on the other hand, the admixture of the $\hat{z}$ component can gap out the line nodes [Fig.~\ref{fig:spectrum_slice_1}(e)].
One may naively expect that the existence of the higher-order term results in only a quantitative change of the spectrum, since the term has the same symmetry as that of the typical Rashba term.
However, the above results indicate that the higher-order effect is qualitatively crucial for the feasibility of supercurrent-induced Weyl superconductivity, which is a surprising point.

In addition, we briefly comment on the difference of the present study from our previous theory~\cite{Takasan2021_arXiv}.
The energy gap arising from the cooperating effect of the supercurrent and the ASOC has already been discussed in 2D $D+p$-wave superconductors~\cite{Takasan2021_arXiv}.
In the 2D model, however, we need to somehow consider an artificial perturbation, i.e., an ASOC term with \textit{different} symmetry from the Rashba term, for the gap opening.
On the other hand, as we explained in Sect.~\ref{sec:model}, the higher-order ASOC term in the 3D model possesses the \textit{same} symmetry as that of the Rashba term, and therefore is naturally introduced without any perturbation.

\subsection{The case of out-of-plane supercurrent}
\label{sec:energy_spectrum_out}
Next, let us consider more general directions of the supercurrent, namely, $q_z \neq 0$.
We find that the quasiparticle energy spectrum under the out-of-plane current is similar to that in the in-plane case, as long as the supercurrent is sufficiently small.
For $|\bm{q}| = 0.1$, indeed, there appear point nodes on the $k_z = 0, \pi$ planes even when the supercurrent has a nonzero $z$-component [see Figs.~\ref{fig:spectrum_slice_4}(a)--\ref{fig:spectrum_slice_4}(c) in Appendix].
The reason can be understood as follows.
As we showed in the previous subsection, the energy gap arising from the supercurrent is given by Eq.~\eqref{eq:gap_by_current}, in which the numerator is calculated as
\begin{equation}
 [\bm{g}'_{\bm{q}} \cdot (\bm{g} \times \bm{d})]_{\bm{k} = \bm{k}_0} = 2 \alpha_1 \alpha_2 \Delta_p (q_x \mp q_y) \sin^5 k_{x0} \sin k_{z0},
 \label{eq:gq_g_d}
\end{equation}
on the node of $\bm{k}_0 = (k_{x0}, \pm k_{x0}, k_{z0})$.
Equations~\eqref{eq:gap_by_current} and \eqref{eq:gq_g_d} indicate that the current-induced gap is independent of $q_z$.
Furthermore, since Eq.~\eqref{eq:gq_g_d} is proportional to $\sin k_{z0}$, the point nodes emerge on the $k_z = 0, \pi$ planes, at least within the perturbation theory.
The fixing of the point nodes on the two $k_z = 0, \pi$ planes is definitely guaranteed under the in-plane supercurrent ($q_z = 0$) by the topological protection (see Sect.~\ref{sec:topology_Berry}), whereas it is not under the out-of-plane current ($q_z \neq 0$).

\begin{figure}[tbp]
 \centering
 \includegraphics[width=.9\linewidth, pagebox=artbox]{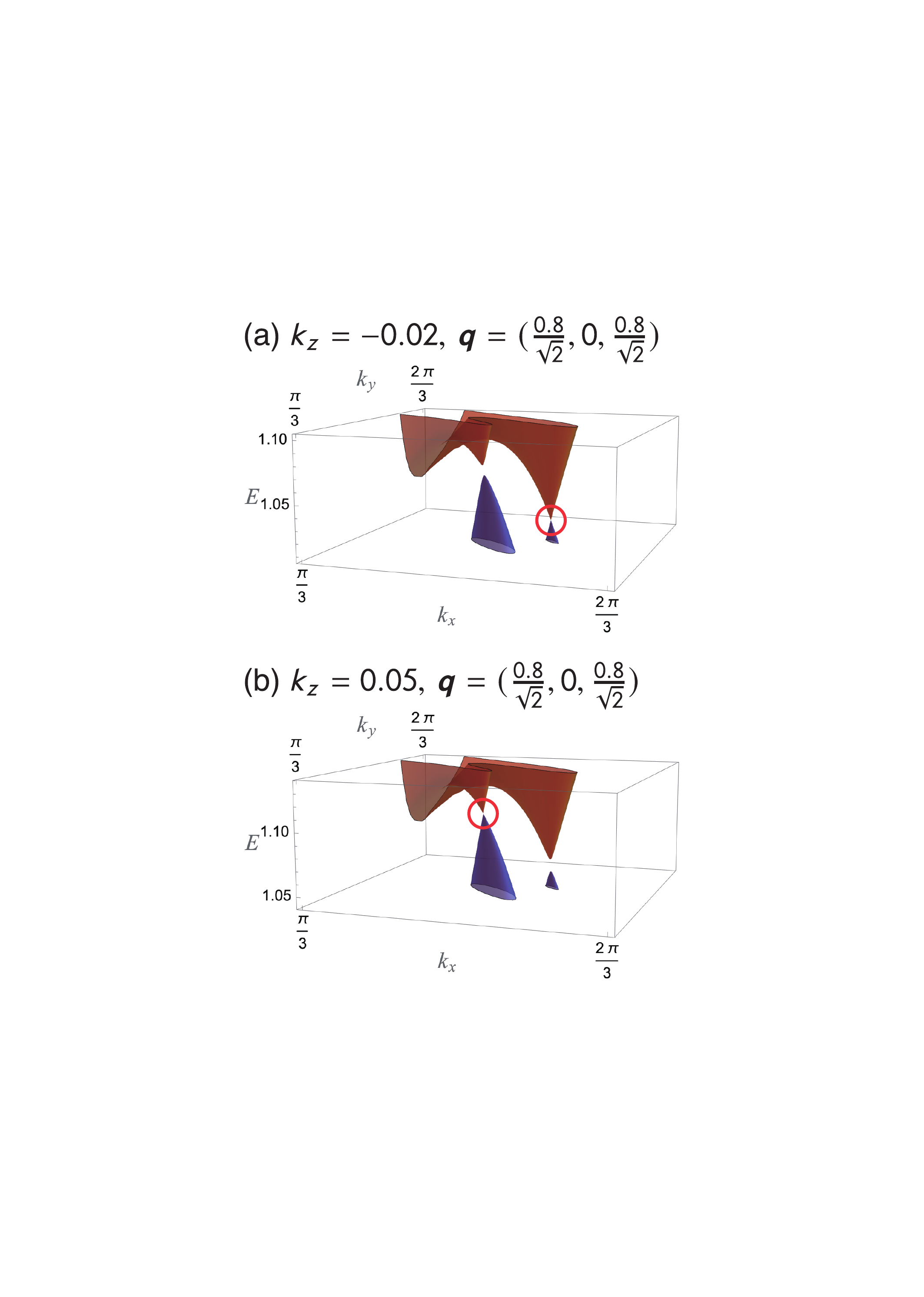}
 \caption{(Color online) Quasiparticle energy spectra on (a) $k_z = -0.02$ and (b) $k_z = 0.05$ planes for $\bm{q} = (\frac{0.8}{\sqrt{2}}, 0, \frac{0.8}{\sqrt{2}})$. The locations of the point nodes deviate from the $k_z = 0$ plane for the large out-of-plane COM momentum (see the red open circles).}
 \label{fig:spectrum_slice_3}
\end{figure}

According to the above discussions, it is expected that the nodes in the out-of-plane-current case can move away from the planes when higher-order effects beyond the perturbation theory are taken into account.
Indeed, we confirm the movement of the nodes by considering the larger out-of-plane supercurrent with $|\bm{q}| = 0.8$ [Figs.~\ref{fig:spectrum_slice_3}(a) and \ref{fig:spectrum_slice_3}(b)].
As we mentioned in Introduction, the tunable positions of the Weyl nodes by the external field are an uncommon feature in the study of Weyl superconductors except UPt$_3$.
Although the huge supercurrent assumed in Figs.~\ref{fig:spectrum_slice_3}(a) and \ref{fig:spectrum_slice_3}(b) would not be realistic due to the existence of a critical current, such a large COM momentum of the Cooper pairs may be feasible in pair-density-wave superconductors~\cite{Agterberg2020_review} as we commented in our previous study~\cite{Takasan2021_arXiv}.

We here note that the current-induced point nodes do not appear for the special $[001]$ direction of the supercurrent.
When $\bm{q} \parallel \hat{z}$, all the line nodes are not gapped out since both $M_{[110]}$ and $M_{[1{-1}0]}$ symmetries are preserved.
We actually confirm that the line nodes remain gapless under the $z$-direction supercurrent [see Figs.~\ref{fig:spectrum_slice_4}(d)--\ref{fig:spectrum_slice_4}(f) in Appendix], which can be also understood in Eq.~\eqref{eq:gq_g_d}.

\section{Topological Aspects of Superconducting Nodes}
\label{sec:topology}
In the previous section, we have shown that the line-nodal structure of the $D+p$-wave superconductor changes to the point-nodal one when the supercurrent is applied.
In this section, we analyze two topological indices characterizing the point nodes: a Weyl charge (Sect.~\ref{sec:topology_Weyl}) and a quantized Berry phase (Sect.~\ref{sec:topology_Berry}).

\subsection{Supercurrent-induced Weyl superconductivity}
\label{sec:topology_Weyl}

\subsubsection{Supercurrent-direction-dependent Weyl charges}
\begin{figure*}
 \centering
 \includegraphics[width=\linewidth, pagebox=artbox]{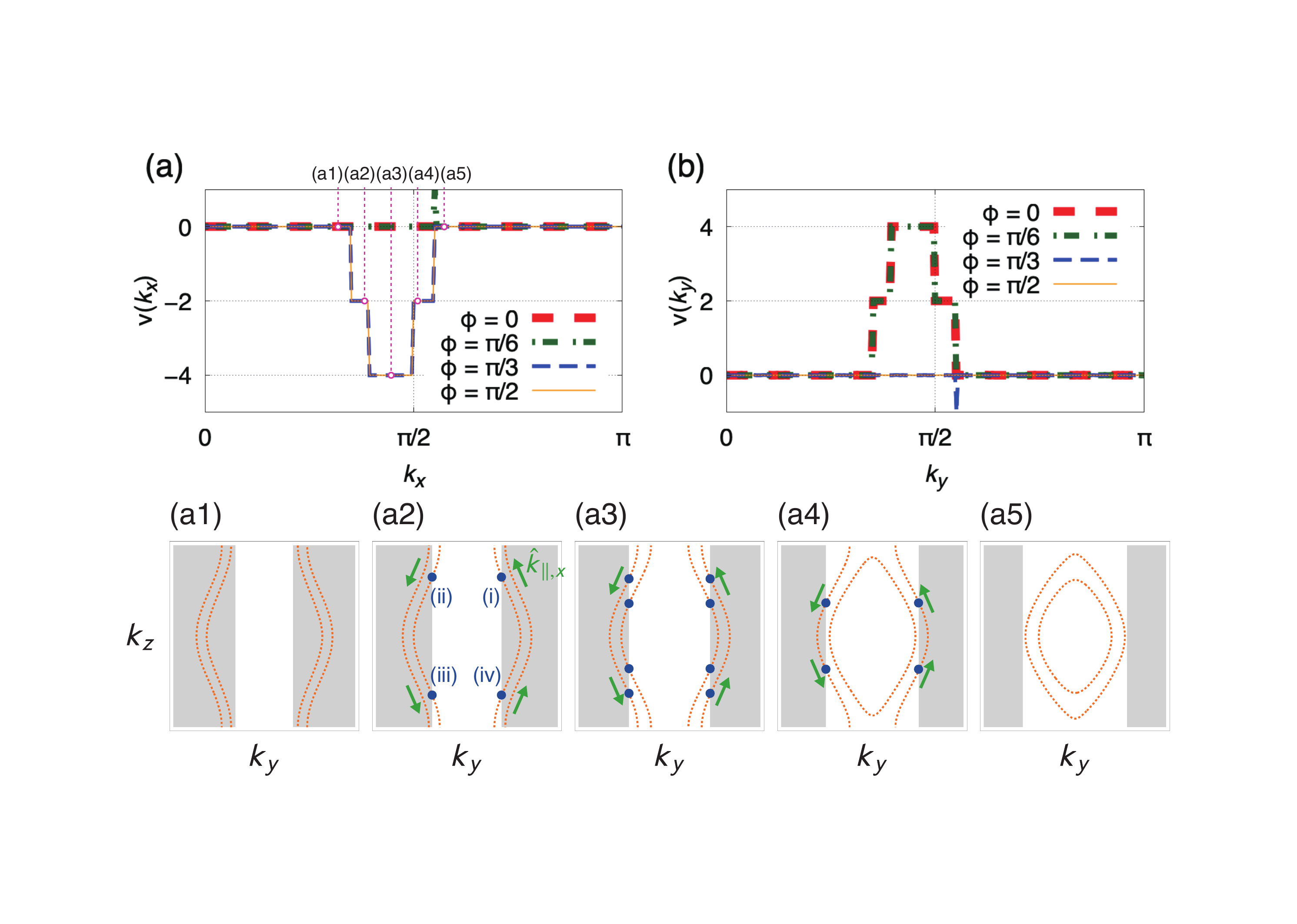}
 \caption{(Color online) (a, b) Chern numbers $\nu(k_x)$ and $\nu(k_y)$ for the in-plane supercurrent with $\bm{q} = 0.1 (\cos\phi, \sin\phi, 0)$. (a1--a5) Normal-state Fermi surfaces (orange dotted lines) and gapless points of the $D+p$-wave superconductivity (blue points) for $\bm{q} = \bm{0}$. Each figure shows a $k_y$--$k_z$ plane, where $k_x$ is fixed on a value depicted in (a). The green arrows show the direction of $\hat{k}_{\parallel, x}$ defined by Eq.~\eqref{eq:kpara}. The shaded regions indicate $\psi(\bm{k}) > 0$.}
 \label{fig:chern_qin}
\end{figure*}

We now clarify that the point nodes on $k_z = 0, \pi$ [Figs.~\ref{fig:spectrum_slice_1}(d) and \ref{fig:spectrum_slice_1}(f)] are identified as topological Weyl nodes, and that the Weyl charges depend on the direction of the supercurrent.
The model Hamiltonian [Eq.~\eqref{eq:Hamiltonian}] is regarded as a superconductor in Altland--Zirnbauer (AZ) class D~\cite{Zirnbauer1996, Altland1997, Schnyder2008, Kitaev2009, Ryu2010}, since time-reversal symmetry is broken under the finite supercurrent.
Thus a topological Weyl charge can be defined as a monopole of Berry flux:
\begin{equation}
 C_j = \frac{1}{2\pi} \iint_S \mathrm{d}\bm{S} \cdot \bm{F}(\bm{k}),
\end{equation}
where $S$ is a closed surface surrounding an isolated point node labeled by $j$.
The Berry flux is given by
\begin{equation}
 \bm{F}(\bm{k}) = \frac{1}{i} \sum_{n=1,2} \nabla_{\bm{k}} \times \Braket{u_n(\bm{k}) | \nabla_{\bm{k}} | u_n(\bm{k})},
\end{equation}
where $\Ket{u_n(\bm{k})}$ is a wavefunction of Bogoliubov quasiparticles with a band index $n$ ($= 1, \dots, 4$).
Note that the summation is taken over the two lowest energy eigenvalues ($n = 1, 2$) for any momentum on $S$.
We identify Weyl nodes by calculating $k_i$-dependent Chern numbers~\cite{Thouless1982, Kohmoto1985, Fukui2005},
\begin{subequations}
 \begin{align}
  \nu(k_x) &= \frac{1}{2\pi} \int_{-\pi}^{\pi} \mathrm{d}k_y \int_{-\pi}^{\pi} \mathrm{d}k_z \, F_x(\bm{k}),
  \label{eq:Chern_x} \\
  \nu(k_y) &= \frac{1}{2\pi} \int_{-\pi}^{\pi} \mathrm{d}k_z \int_{-\pi}^{\pi} \mathrm{d}k_x \, F_y(\bm{k}).
  \label{eq:Chern_y}
 \end{align}
\end{subequations}
When the Chern number $\nu(k_x)$ jumps at $k_x = k_{x0}$, its value is equal to the sum of Weyl charges at $k_{x0}$:
\begin{equation}
 \nu(k_{x0} + 0) - \nu(k_{x0} - 0) = \sum_{j \in \{\text{nodes \ at} \ k_{x0}\}} C_j.
 \label{eq:Weyl_charge_x}
\end{equation}
Similarly, a jump of $\nu(k_y)$ at $k_y = k_{y0}$ is given by the sum of Weyl charges at $k_{y0}$,
\begin{equation}
 \nu(k_{y0} + 0) - \nu(k_{y0} - 0) = \sum_{j \in \{\text{nodes \ at} \ k_{y0}\}} C_j.
 \label{eq:Weyl_charge_y}
\end{equation}

\begin{figure}[tbp]
 \centering
 \includegraphics[width=\linewidth, pagebox=artbox]{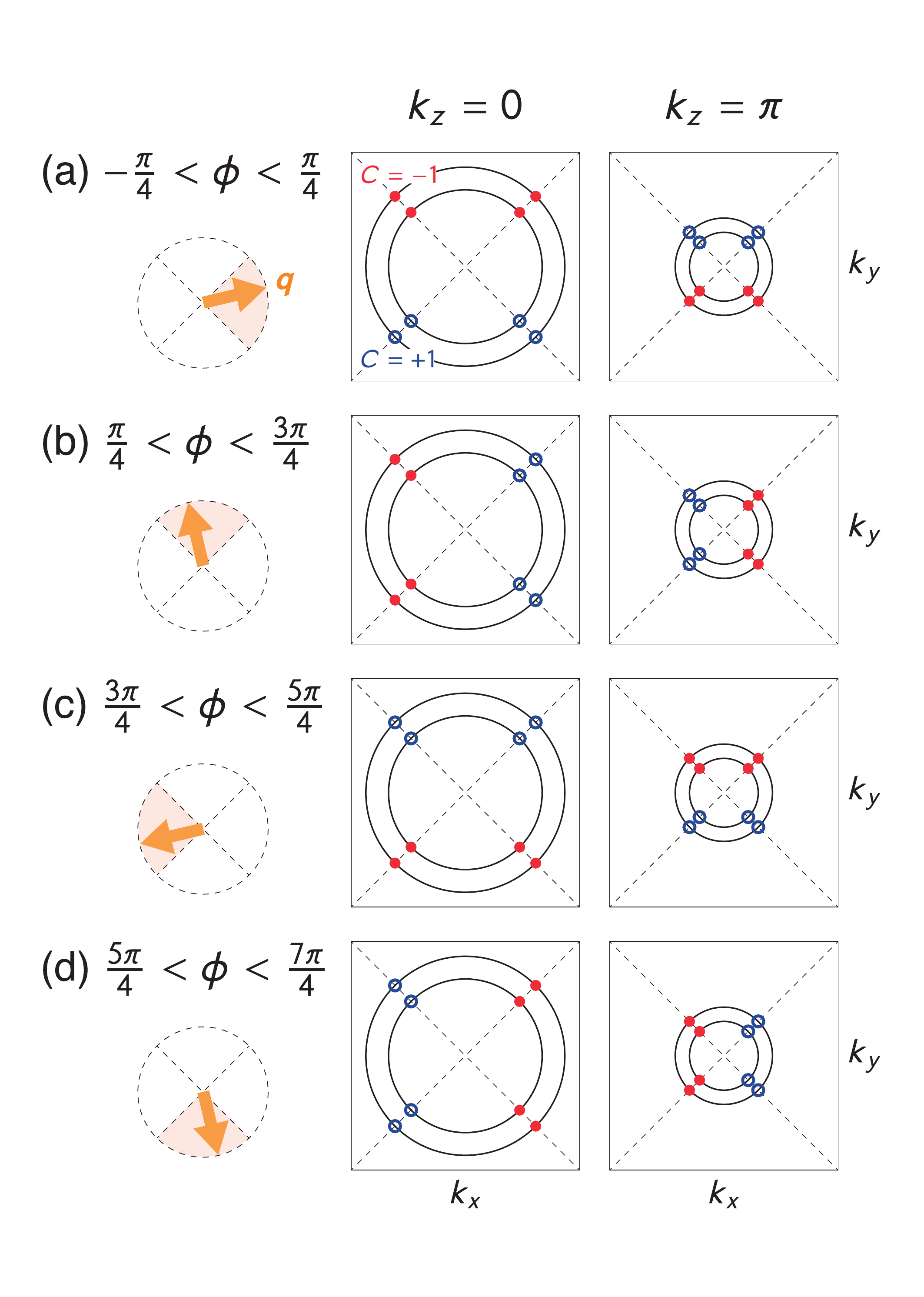}
 \caption{(Color online) Schematic pictures of Weyl nodes induced by the in-plane supercurrent with $\bm{q} = q (\cos\phi, \sin\phi, 0)$ on the $k_z = 0, \pi$ planes, for (a) $-\pi/4 < \phi < \pi/4$, (b) $\pi/4 < \phi < 3\pi/4$, (c) $3\pi/4 < \phi < 5\pi/4$, and (d) $5\pi/4 < \phi < 7\pi/4$. The orange arrows show the direction of the COM momentum $\bm{q}$. The black solid lines indicate the Fermi surfaces in the normal state, and blue open and red solid circles represent Weyl nodes with $C_j = +1$ and $-1$, respectively.}
 \label{fig:weyl_nodes}
\end{figure}

Figure~\ref{fig:chern_qin}(a) shows the Chern number $\nu(k_x)$ [Eq.~\eqref{eq:Chern_x}] for the COM momentum $\bm{q} = 0.1 (\cos\phi, \sin\phi, 0)$.
When the azimuth $\phi$ is equal to $0$ and $\pi/6$, $\nu(k_x)$ is zero for almost all $k_x$, whereas plateaus of $\nu(k_x) < 0$ appear for $\phi = \pi/3$ and $\pi/2$~%
\footnote{In Fig.~\ref{fig:chern_qin}(a), a sharp peak with $\nu(k_x) = 1$ is shown for $\phi = \pi/6$. This is not a numerical error. The reason is that the two rightmost Weyl nodes on $k_z = 0$ in Fig.~\ref{fig:weyl_nodes}(a) are located at slightly different $k_x$ values for the ``halfway'' angle $\phi = \pi/6$, whereas the two nodes are aligned at the same $k_x$ for $\phi = 0$. The same discussion is applicable to a sharp peak with $\nu(k_y) = -1$ for $\phi = \pi/3$ in Fig.~\ref{fig:chern_qin}(b).}.
On the other hand, the $k_y$-dependent Chern number has plateaus of $\nu(k_y) > 0$ for $\phi = 0$ and $\pi/6$, and is zero for $\phi = \pi/3$ and $\pi/2$ [Fig.~\ref{fig:chern_qin}(b)].
From Eqs.~\eqref{eq:Chern_x} and \eqref{eq:Chern_y}, the results indicate that the point nodes on $k_z = 0, \pi$ [Figs.~\ref{fig:spectrum_slice_1}(d) and \ref{fig:spectrum_slice_1}(f)] are topologically protected Weyl points, whose Weyl charges depend on the direction of the supercurrent, as illustrated in Figs.~\ref{fig:weyl_nodes}(a) and \ref{fig:weyl_nodes}(b).
For $-\pi/4 < \phi < \pi/4$, the sign of the Weyl charges on $k_z = 0$ is negative (positive) for $k_y > 0$ ($k_y < 0$), while the sign is exchanged on $k_z = \pi$ [see Fig.~\ref{fig:weyl_nodes}(a)].
However, the Weyl charges in the $[110]$ direction change their sign for $\pi/4 < \phi < 3\pi/4$ [Fig.~\ref{fig:weyl_nodes}(b)].
The sign change is considered as a kind of ``topological phase transition.''
Indeed, when the azimuth is on the ``phase boundary,'' namely $\phi = \pi/4$, the Weyl nodes in the $[110]$ direction are ill-defined since the line nodes appear in the direction, as we explained in Sect.~\ref{sec:energy_spectrum_in}.
For the other regimes of $\phi$, we can similarly discuss the sign change of the Weyl charges [see Figs.~\ref{fig:weyl_nodes}(c) and \ref{fig:weyl_nodes}(d)].

The $\phi$ dependence of the Weyl charge can be understood as follows.
First, we carry out a unitary transformation of the BdG Hamiltonian [Eq.~\eqref{eq:Hamiltonian}] by a unitary matrix
\begin{equation}
 U =
 \begin{bmatrix}
  \sigma_0 & 0 \\
  0 & -\ii\sigma_y
 \end{bmatrix}.
\end{equation}
Then the Hamiltonian takes the following form:
\begin{align}
 \tilde{H}_{\text{BdG}}(\bm{k}; \bm{q}) &:= U H_{\text{BdG}}(\bm{k}; \bm{q}) U^\dagger \notag \\
 &=
 \begin{bmatrix}
  H_{\text{N}}(\bm{k} + \bm{q}) & \tilde{\Delta}(\bm{k}) \\
  \tilde{\Delta}^\dagger(\bm{k}) & - (\ii\sigma_y) H_{\text{N}}^*(-\bm{k} + \bm{q}) (-\ii\sigma_y)
 \end{bmatrix},
\end{align}
where $\tilde{\Delta}(\bm{k}) = \psi(\bm{k}) \sigma_0 + \bm{d}(\bm{k}) \cdot \bm{\sigma}$.
By using the above representation, let us consider the topological property of the BdG Hamiltonian under the small supercurrent.
For the purpose, we expand the Hamiltonian up to the first order of $\bm{q}$,
\begin{align}
 \tilde{H}_{\text{BdG}}(\bm{k}; \bm{q}) &= H_{\text{N}}(\bm{k}) \tau_z + \tilde{\Delta}(\bm{k}) \tau_x \notag \\
 &\quad + \xi'_{\bm{q}}(\bm{k}) \sigma_0 \tau_0 + \bm{g}'_{\bm{q}}(\bm{k}) \cdot \bm{\sigma} \tau_0 + \mathcal{O}(q^2),
 \label{eq:Hamiltonian_expansion}
\end{align}
where $\xi'_{\bm{q}}(\bm{k}) = (\bm{q} \cdot \nabla_{\bm{k}}) \xi(\bm{k})$ and $\bm{\tau} = (\tau_x, \tau_y, \tau_z)$ represents the Pauli matrices in the Nambu space.
The third term proportional to the identity matrix generates the momentum-dependent energy shift known as the Doppler shift~\cite{Matsuda2006}.
Here we can neglect the third term since it does not influence the topological discussion.
Therefore, the leading effect of the supercurrent for the topological Weyl charge is represented by the fourth term in Eq.~\eqref{eq:Hamiltonian_expansion}.

From the above discussion, the topological property of Eq.~\eqref{eq:Hamiltonian_expansion} can be obtained by investigating the following effective Hamiltonian,
\begin{equation}
 H_{\text{N}}(\bm{k}) \tau_z + \tilde{\Delta}(\bm{k}) \tau_x + \bm{g}'_{\bm{q}}(\bm{k}) \cdot \bm{\sigma} \tau_0,
 \label{eq:Hamiltonian_effective}
\end{equation}
which has the same structure as the model of an inversion-symmetry-breaking superconductor under a Zeeman field $\bm{h}$:
\begin{equation}
 H_{\text{N}}(\bm{k}) \tau_z + \tilde{\Delta}(\bm{k}) \tau_x - \bm{h} \cdot \bm{\sigma} \tau_0.
\end{equation}
The topological nature of such models has been discussed in previous theoretical studies~\cite{Daido2016, Yanase2022_review}.
In particular, when the spin-singlet ($d$-wave) component of the order parameter is dominant, the Chern numbers [Eqs.~\eqref{eq:Chern_x} and \eqref{eq:Chern_y}] of Eq.~\eqref{eq:Hamiltonian_effective} are evaluated by~%
\footnote{The formula can be derived supposing that the conditions
\begin{equation*}
|\psi| \ll |\bm{g}|, \, |\bm{d}| \ll |\bm{g}|, \, |\bm{g}'_{\bm{q}}| \ll |\bm{g}|,
\end{equation*}
are satisfied around $\bm{k}_0$~\cite{Daido2016, Yanase2022_review}. These assumptions are valid unless the gap nodes are located at zeros of the $g$-vector or the applied supercurrent is large.}
\begin{equation}
 \nu(k_i) = \sum_{\bm{k}_0} \frac{1}{2} \sgn\left[ \frac{\partial\psi/\partial k_{\parallel, i}}{-\bm{g}'_{\bm{q}} \cdot (\bm{g} \times \bm{d})} \right]_{\bm{k} = \bm{k}_0} \quad (i = x, y),
 \label{eq:Chern_formula}
\end{equation}
where the summation is taken over the nodal points on the $k_i$ slice in the $\bm{q} \to \bm{0}$ limit.
$\hat{k}_{\parallel, i}$ is the direction along the Fermi surface:
\begin{equation}
 \hat{k}_{\parallel, x} := \hat{x} \times \hat{k}_{\perp}, \,
 \hat{k}_{\parallel, y} := \hat{y} \times \hat{k}_{\perp}, \,
 \hat{k}_{\perp} := \frac{\nabla_{\bm{k}} E_\pm(\bm{k})}{|\nabla_{\bm{k}} E_\pm(\bm{k})|},
 \label{eq:kpara}
\end{equation}
where $E_\pm(\bm{k}) := \xi(\bm{k}) \pm |\bm{g}(\bm{k})|$ is the normal energy band with the helicity $\pm$.
In our model, the denominator in the sign function of Eq.~\eqref{eq:Chern_formula} is given by Eq.~\eqref{eq:gq_g_d}.
The equation includes the COM momentum in the form of $(q_x \mp q_y)$, which causes the $\phi$ dependence of the Weyl charge.

To check the validity of Eq.~\eqref{eq:Chern_formula}, let us reproduce the numerical results of the Chern number in Fig.~\ref{fig:chern_qin}(a) by using the formula.
Figures~\ref{fig:chern_qin}(a1)--\ref{fig:chern_qin}(a5) represent normal-state Fermi surfaces on $k_x$-fixed slices depicted in Fig.~\ref{fig:chern_qin}(a).
In Fig.~\ref{fig:chern_qin}(a2), for example, four gapless points [the blue points labeled by (i)--(iv)] appear on the Fermi surfaces in the superconducting state with $\bm{q} = \bm{0}$.
When the finite supercurrent is applied, the four nodes are gapped out~%
\footnote{In order to gap out the nodes, the COM momentum $\bm{q}$ should not be parallel any of the $[110]$, $[1{-1}0]$, and $[001]$ directions.},
each of which contributes to the Chern number on the slice.
By using Fig.~\ref{fig:chern_qin}(a2) and Eqs.~\eqref{eq:parameter} and \eqref{eq:gq_g_d}, we can easily derive the following relations:
\begin{equation}
 \begin{array}{ccc} \hline\hline
  \text{Node} & \sgn[\partial\psi/\partial k_{\parallel, x}]_{\bm{k} = \bm{k}_0} & \sgn[-\bm{g}'_{\bm{q}} \cdot (\bm{g} \times \bm{d})]_{\bm{k} = \bm{k}_0} \\ \hline
  \text{(i)} & - & - \sgn(q_x - q_y) \\
  \text{(ii)} & + & - \sgn(q_x + q_y) \\
  \text{(iii)} & - & + \sgn(q_x + q_y) \\
  \text{(iv)} & + & + \sgn(q_x - q_y) \\ \hline\hline
 \end{array}
\end{equation}
From the formula~\eqref{eq:Chern_formula}, therefore, the Chern number is given by
\begin{equation}
 \nu(k_x) =
 \begin{cases}
  0,  & (q_x + q_y)(q_x - q_y) > 0, \\
  -2, & q_x + q_y > 0, \, q_x - q_y < 0, \\
  +2, & q_x + q_y < 0, \, q_x - q_y > 0,
 \end{cases}
\end{equation}
which is consistent with the results in Fig.~\ref{fig:chern_qin}(a).
Similarly, we can confirm the validity of Eq.~\eqref{eq:Chern_formula} on the other slices.

\subsubsection{Surface arc states}
\label{sec:topology_Weyl_arc}
\begin{figure}[tbp]
 \centering
 \includegraphics[width=\linewidth, pagebox=artbox]{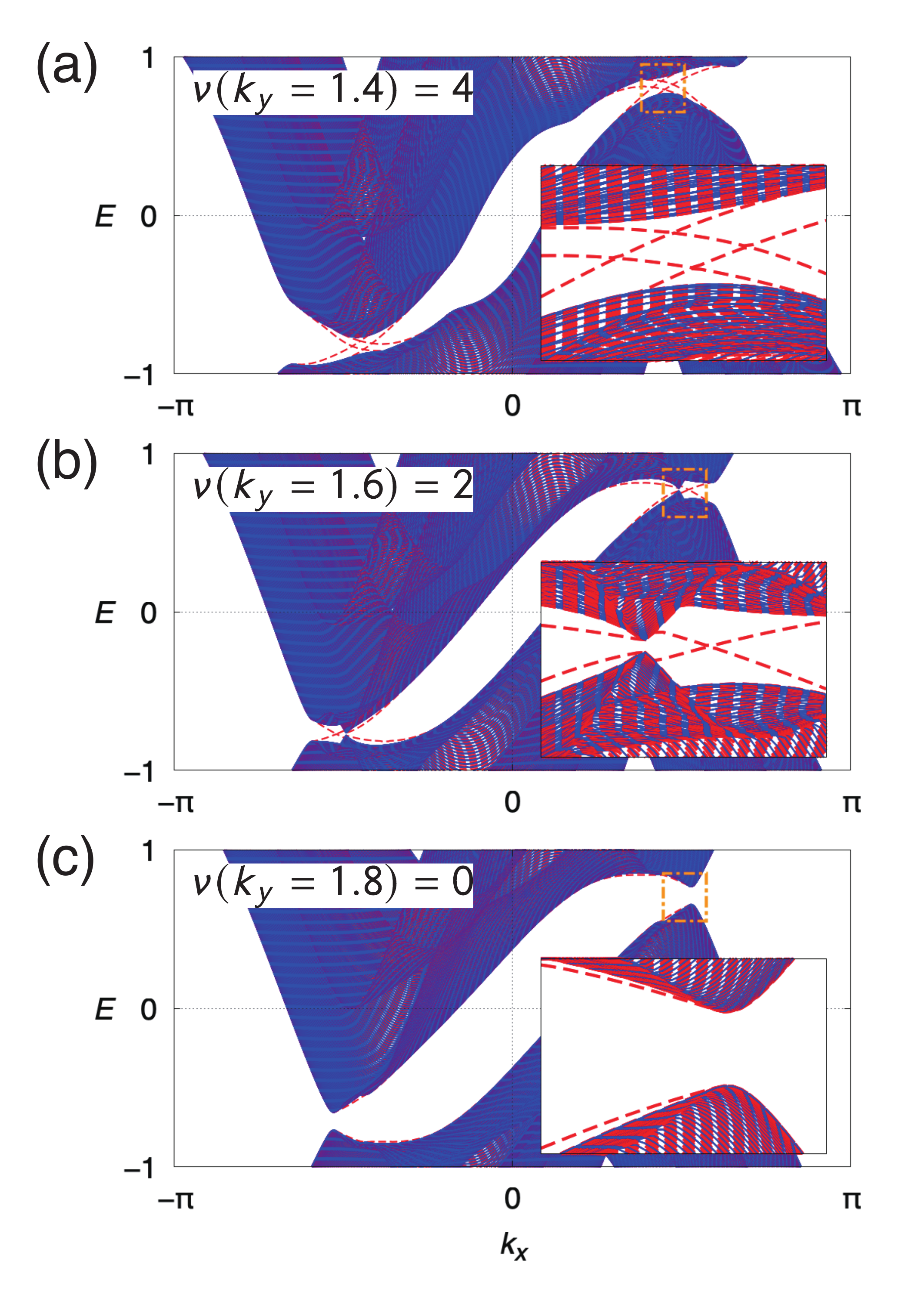}
 \caption{(Color online) Energy spectra of Bogoliubov quasiparticles for the COM momentum $\bm{q} = (0.4, 0, 0)$ at (a) $k_y = 1.4$, (b) $k_y = 1.6$, and (c) $k_y = 1.8$. The blue solid and red dashed lines represent the BdG spectrum under the PBC and OBC along the $z$ axis, respectively. The inset in each figure indicates the enlarged energy spectrum in the orange dashed-dotted box.}
 \label{fig:spectrum_obc}
\end{figure}

We show that the Weyl nodes emerging in the presence of the supercurrent host surface arc states under open boundary conditions (OBC).
Figures~\ref{fig:spectrum_obc}(a)--\ref{fig:spectrum_obc}(c) show quasiparticle energy spectra under the $x$-direction supercurrent on the slices with $k_y = 1.4$, $1.6$, and $1.8$.
The blue solid lines are the spectra obtained by considering periodic boundary conditions (PBC) for all directions.
On the other hand, the red dashed lines represent the results calculated in the presence of the OBC along the $z$ axis and the PBC along the $x$ and $y$ directions.
In Figs.~\ref{fig:spectrum_obc}(a) and \ref{fig:spectrum_obc}(b), gapless surface states obviously appear in the bulk gap, while they do not in Fig.~\ref{fig:spectrum_obc}(c).
The gapless modes are surface arc states corresponding to the nontrivial Weyl nodes located on the $k_z = 0, \pi$ planes (Fig.~\ref{fig:weyl_nodes}).
Indeed, on the $k_y$ slices of Figs.~\ref{fig:spectrum_obc}(a)--\ref{fig:spectrum_obc}(c), the Chern number $\nu(k_y)$ has the values $4$, $2$, and $0$, respectively [see Fig.~\ref{fig:chern_qin}(b)].
Here note that the surface arc states are in general not around the zero energy because of the Doppler shift in the presence of the supercurrent.

\subsection{Quantized Berry phase on \texorpdfstring{$k_z = 0, \pi$}{kz = 0, pi} for in-plane supercurrent}
\label{sec:topology_Berry}
Let us move on to discussions about another topological index, a quantized Berry phase, which helps us to understand why the supercurrent-induced Weyl nodes are fixed on the $k_z = 0, \pi$ planes for the in-plane supercurrent.

The Berry phase $\gamma_C$ is defined as
\begin{equation}
 \gamma_C = \frac{1}{i} \sum_{n=1,2} \oint_C \mathrm{d}\bm{k} \cdot \Braket{u_n(\bm{k}) | \nabla_{\bm{k}} | u_n(\bm{k})} \pmod{2\pi},
 \label{eq:Berry_phase}
\end{equation}
where $C$ is a 1D closed path in the Brillouin zone.
The bands of $n = 1,2$ are taken as the two lowest energy eigenstates for each momentum.
In the absence of the supercurrent, the chiral symmetry $\Gamma$ ensures the $\mathbb{Z}_2$ quantization of the Berry phase, which coincides with the parity of the 1D winding number in Eq.~\eqref{eq:winding_num}~\cite{Ryu2010, Takasan2021_arXiv}:
\begin{equation}
 \gamma_C = \pi w_C \pmod{2\pi}.
\end{equation}
On the other hand, the finite supercurrent violates the conservation of the time-reversal symmetry $T$ and the chiral symmetry $\Gamma$.
As a result, the winding number is ill-defined, and the quantization of the Berry phase breaks down for a general path $C$.
For a special loop $C$ included in the $k_z = 0$ or $k_z = \pi$ plane, however, the Berry phase is quantized even when the in-plane supercurrent is applied~\cite{Takasan2021_arXiv}.
This is because $TC_{2z}$ symmetry, namely the combination of the twofold rotation symmetry $C_{2z}$ and the time-reversal symmetry $T$, is preserved at any $\bm{k}$ point on the $k_z = 0, \pi$ planes, and squares to $+1$.
The symmetry configuration can be identified as the AI class within a previous AZ$+\mathcal{I}$ classification theory~\cite{Bzdusek2017}, which indicates the presence of $\mathbb{Z}_2$ topology in one dimension, i.e., the quantization of the Berry phase.

The Berry phase helps our understanding of the quasiparticle energy spectra in Figs.~\ref{fig:spectrum_slice_1}(a)--\ref{fig:spectrum_slice_1}(f) as follows.
The $D+p$-wave order parameter results in the existence of the line nodes for $\bm{q} = \bm{0}$ [Figs.~\ref{fig:spectrum_slice_1}(a)--\ref{fig:spectrum_slice_1}(c)], which are protected by the quantized Berry phase [Eq.~\eqref{eq:Berry_phase}] as well as the winding number [Eq.~\eqref{eq:winding_num}].
When the in-plane supercurrent is applied, the nodes on a general momentum are gapped out due to the breakdown of the two topological invariants [Fig.~\ref{fig:spectrum_slice_1}(e)].
On the other hand, the nodes on the $k_z = 0, \pi$ planes remain since the $TC_{2z}$ symmetry keeps the $\mathbb{Z}_2$ quantization of the Berry phase [Figs.~\ref{fig:spectrum_slice_1}(d) and \ref{fig:spectrum_slice_1}(f)].
That is why there appear point nodes fixed on the $k_z = 0, \pi$ planes under the in-plane supercurrent, which are characterized by the Weyl charge as we showed in Sect.~\ref{sec:topology_Weyl}.

\begin{figure}[tbp]
 \centering
 \includegraphics[width=8cm, pagebox=artbox]{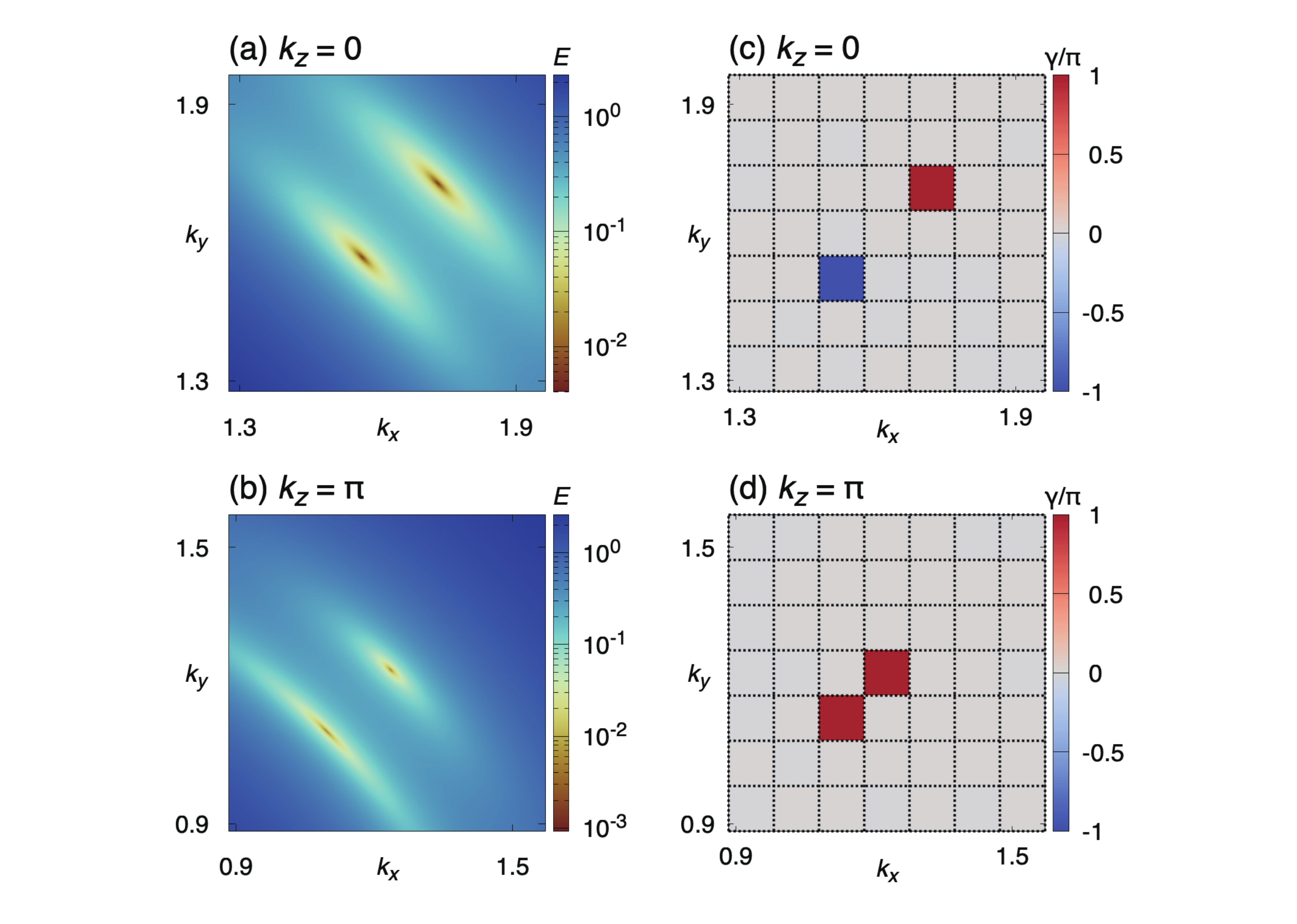}
 \caption{(Color online) Heat maps of (a, b) quasiparticle energy gap [Eq.~\eqref{eq:energy_gap}] and (c, d) quantized Berry phase [Eq.~\eqref{eq:Berry_phase}] for the COM momentum $\bm{q} = (0.1, 0, 0)$. Upper (lower) panels show the maps on the $k_z = 0$ ($k_z = \pi$) plane. The intense spots in (a) and (b) represent the gapless points shown in Figs.~\ref{fig:spectrum_slice_1}(d) and \ref{fig:spectrum_slice_1}(f), respectively. In (c) and (d), the Berry phase on each plaquette is calculated by integration along the perimeter of the plaquette (the dotted lines).}
 \label{fig:spectrum_berry}
\end{figure}

In order to make sure the above discussion, we show the numerical results on the $k_z = 0, \pi$ planes of the quasiparticle energy gap in Figs.~\ref{fig:spectrum_berry}(a) and \ref{fig:spectrum_berry}(b), and the Berry phase in Figs.~\ref{fig:spectrum_berry}(c) and \ref{fig:spectrum_berry}(d), respectively.
The energy gap is defined by 
\begin{equation}
 E_3(\bm{k}) - E_2(\bm{k}),
 \label{eq:energy_gap}
\end{equation}
where $E_n(\bm{k})$ is energy eigenvalues of the BdG Hamiltonian with $n = 1, \dots, 4$ and $E_n(\bm{k}) < E_m(\bm{k})$ for $n < m$.
The Berry phase is calculated using a method for the discretized Brillouin zone~\cite{Hatsugai2006}.
These figures obviously indicate that the Berry phase takes a non-zero quantized value~%
\footnote{Although there seem to be three quantized values of the Berry phase in Fig.~\ref{fig:spectrum_berry}(c), the blue and red plaquettes represent $\gamma = \pi + (-0)$ and $-\pi + (+0)$, respectively, which are the same $\mathbb{Z}_2$ index modulo $2\pi$. The tiny fluctuations $\pm 0$ arise from the numerical error.},
when the path $C$ encloses the point node [the intense spots in Figs.~\ref{fig:spectrum_berry}(a) and \ref{fig:spectrum_berry}(b)].

Finally, we briefly comment on the case of out-of-plane supercurrent.
In this case, the $\mathbb{Z}_2$ quantization of the Berry phase [Eq.~\eqref{eq:Berry_phase}] does not occur since the $TC_{2z}$ symmetry is broken~%
\footnote{We actually confirmed that the Berry phase is not quantized for the out-of-plane supercurrent in numerical calculations.}.
Indeed, we showed that the Weyl nodes can move away from the $k_z = 0, \pi$ planes [see Figs.~\ref{fig:spectrum_slice_3}(a) and \ref{fig:spectrum_slice_3}(b)].
When the supercurrent is sufficiently small, however, the Weyl nodes remain at the almost same position even when the current deviates from the in-plane direction [Figs.~\ref{fig:spectrum_slice_4}(a)--\ref{fig:spectrum_slice_4}(c) in Appendix].
This is explained not by the topological origin, but by the perturbation theory [Eqs.~\eqref{eq:gap_by_current} and \eqref{eq:gq_g_d}].

\section{Summary and Discussion}
\label{sec:summary}
In this paper, we investigated the 3D tight-binding model of the noncentrosymmetric $D+p$-wave superconductor under the finite supercurrent, by considering the uniform COM momentum $2\bm{q}$.
We found that the line nodes in the parent superconductor are changed into the point nodes in the finite-current state, which are characterized by the topological Weyl charge.
The positions as well as the Weyl charges of the point nodes depend on the direction of the supercurrent.
Furthermore, we clarified that the Weyl points are located on the high-symmetry planes $k_z = 0$ and $\pi$, when the in-plane supercurrent is applied.
The pinning of the nodes is topologically protected by the $\mathbb{Z}_2$-quantized Berry phase because the combined $TC_{2z}$ symmetry is preserved.
We also elucidated that the higher-order effect in the ASOC is crucial for the above phenomena.

We briefly discuss experimental setups to observe the supercurrent-induced Weyl superconductivity.
The key ingredients for our proposal are the line nodes in the parent superconductor and the ASOC arising from inversion symmetry breaking.
We thus consider that heavy-fermion superconductors CeRhSi$_3$~\cite{Kimura2005, Terashima2007, Muro2007, Kimura2007_PRL, Kimura2007_JPSJ, Tada2008_PRL, Tada2008_JPSJ, Tada2010, Landaeta2018, Ivanov2021} and CeIrSi$_3$~\cite{Sugitani2006, Okuda2007, Settai2008, Mukuda2008, Mukuda2010, Tada2008_PRL, Tada2008_JPSJ, Tada2010, Landaeta2018, Ivanov2021} are good candidates.
Both materials have a noncentrosymmetric tetragonal lattice structure and experience pressure-induced superconductivity with line nodes, while they show an itinerant antiferromagnetic order at ambient pressure.
Another promising candidate is few-layered cuprates fabricated on substrates.
Although in Ref.~\cite{Takasan2021_arXiv} we considered cuprate thin films as a platform of current-induced topological phase transition in 2D superconductors, the 3D-like properties discussed in this paper may appear in the thicker systems.
Furthermore, a small sample is necessary to realize the uniform supercurrent considered in our theory.
In the above candidate superconductors, the London penetration depth $\lambda$ or the Pearl length $\Lambda = 2\lambda^2 / d$ ($d$: the sample thickness) are estimated to 0.1--1{\textmu}m~\cite{Bauer-Sigrist, Pearl1964, Clem2011}.
Recent developments in nanopatterning techniques for superconductors~\cite{Nawaz2013, Li2013, Sun2020} would enable us to make a sample smaller than the length scales.

As we discussed in our previous paper~\cite{Takasan2021_arXiv}, our theory is applicable to Fulde--Ferrell (FF) superconductivity with a single COM momentum of the Cooper pairs~\cite{FF}, and pair-density-wave superconductivity~\cite{Agterberg2020_review}.
In this direction, superlattices of $d$-wave superconductor CeCoIn$_5$ could be a good candidate, where helical or stripe superconductivity may be realized in the high-magnetic-field phase~\cite{Naritsuka2017}.
Another promising possibility is FF superconductivity coexisting with an odd-parity magnetic multipole order~\cite{Sumita2016, Sumita2017}.
For example, the previous theoretical study has suggested that electron-doped Sr$_2$IrO$_4$ in the magnetic quadrupole state has the potential to realize the finite-COM-momentum pairing~\cite{Sumita2017}, whereas bulk superconductivity has not been observed in this material.
However, we expect that other good candidates will be discovered, since the recent symmetry-based approach has identified more than 110 odd-parity magnetic multipole materials~\cite{Watanabe2018}.

Finally, we also comment on experimental methods.
One promising way to observe the current-induced Weyl superconductivity is to measure the density of states that reflects the change of the node structure.
It is possible in various experimental techniques such as optical spectroscopy or scanning tunneling spectroscopy.
More direct evidence for the non-trivial topology is the observation of the surface arc states discussed in Sect.~\ref{sec:topology_Weyl_arc}.
In particular, as shown in Fig.~\ref{fig:spectrum_obc}, the arc states can be below the Fermi energy due to the Doppler shift.
This can be detected in principle using the ARPES, which probes the quasiparticle spectrum of the filled states, and it can be a clear signature of the Weyl superconductivity.
Although we admit that the current-induced gap is not so large and the experimental detection is challenging, many theoretical studies have suggested topological superconductivity with a finite COM momentum~\cite{Romito2012, Lesser2021, Zhang2013, Qu2013, Nissinen2017, Volovik2018, Ying2018, Ying2019, Hu2019, Dmytruk2019, Papaj2021, Volkov2020_arXiv, Takasan2021_arXiv}.
Therefore, we expect that further promising experimental setups are proposed and such a topological phase is observed in future works.

%%%%%%%%%% Acknowledgments %%%%%%%%%%
\begin{acknowledgments}
 The authors are grateful to Youichi Yanase and Akito Daido for fruitful discussions and comments.
 S.S. was supported by JST CREST Grant No. JPMJCR19T2.
 K.T. was supported by the U.S. Department of Energy (DOE), Office of Science, Basic Energy Sciences (BES), under Contract No. AC02-05CH11231 within the Ultrafast Materials Science Program (KC2203).
\end{acknowledgments}

\appendix
\section{Numerical Results for Out-of-plane Supercurrent}

\begin{figure*}[tbp]
 \centering
 \includegraphics[width=\linewidth, pagebox=artbox]{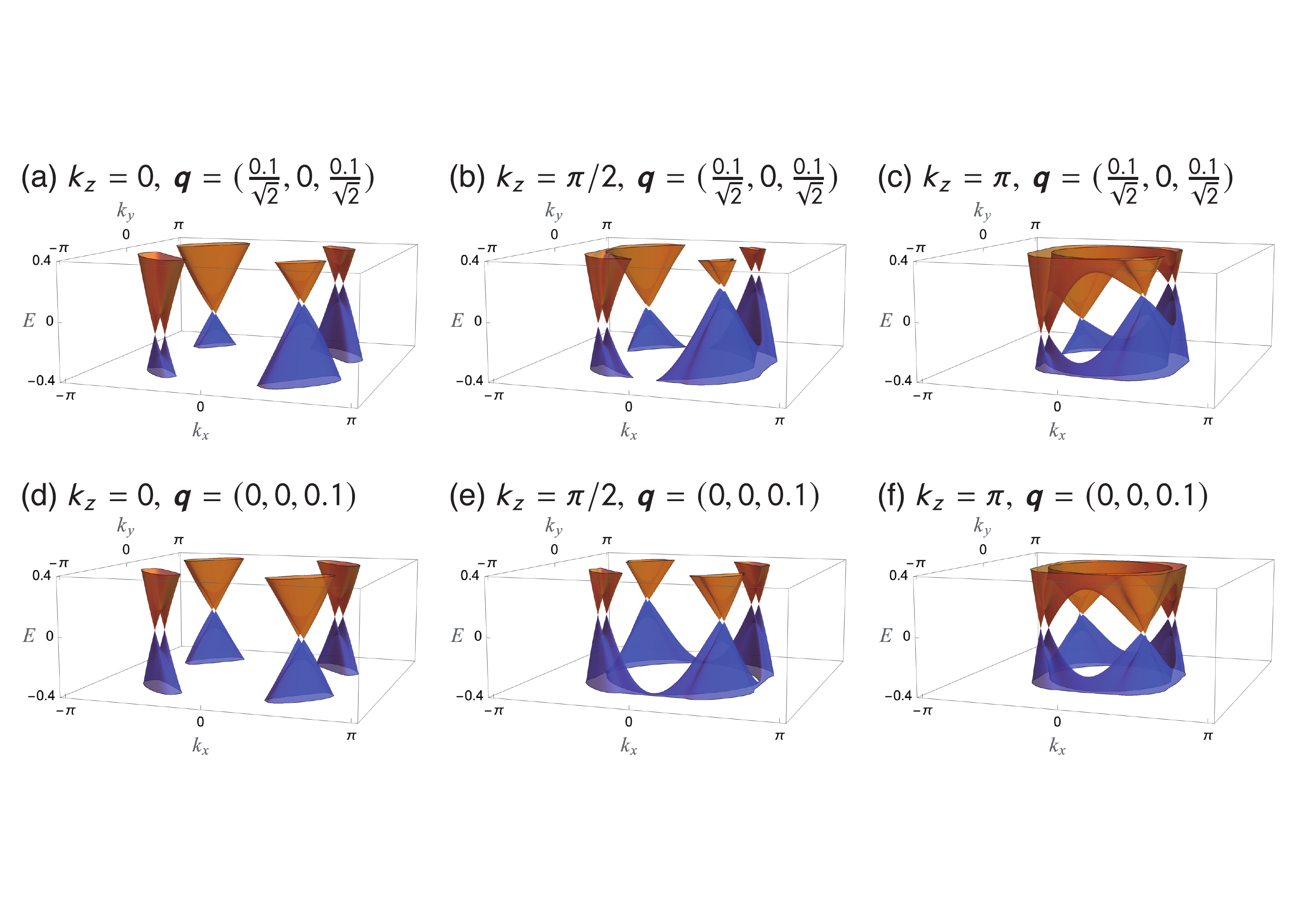}
 \caption{(Color online) Quasiparticle energy spectra on (a, d) $k_z = 0$, (b, e) $k_z = \pi/2$, and (c, f) $k_z = \pi$ planes. The COM momentum $\bm{q}$ of the Cooper pairs is set to be $(\frac{0.1}{\sqrt{2}}, 0, \frac{0.1}{\sqrt{2}})$ in the upper panels (a--c), while it is $(0, 0, 0.1)$ in the lower panels (d--f).}
 \label{fig:spectrum_slice_4}
\end{figure*}

We show the Bogoliubov quasiparticle energy spectrum in the case of the out-of-plane supercurrent (Sect.~\ref{sec:energy_spectrum_out}).
Figures~\ref{fig:spectrum_slice_4}(a)--\ref{fig:spectrum_slice_4}(c) and \ref{fig:spectrum_slice_4}(d)--\ref{fig:spectrum_slice_4}(f) represent the numerical results for $\bm{q} = (\frac{0.1}{\sqrt{2}}, 0, \frac{0.1}{\sqrt{2}})$ and $\bm{q} = (0, 0, 0.1)$, respectively.

% \bibliography{paper}

%

\end{document}